# Improved second harmonic performance in periodically poled LNOI waveguides through engineering of lateral leakage

ANDREAS BOES,[1,2,*] LIN CHANG,[2] MARKUS KNOERZER,[1]
THACH G. NGUYEN,[1] JON D. PETERS,[2] JOHN E. BOWERS,[2] AND
ARNAN MITCHELL[1]

[1]*School of Engineering, RMIT University, Melbourne, VIC 3000, Australia*
[2]*Department of Electrical and Computer Engineering, University of California, Santa Barbara, CA 93106, USA*
*andreas.boes@rmit.edu.au*

**Abstract:** In this contribution we investigate the impact of lateral leakage for linear and nonlinear optical waveguides in lithium niobate on insulator (LNOI). Silicon nitride (SiN) loaded and direct patterned lithium niobate cross-sections are investigated. We show that lateral leakage can take place for the TE mode in LNOI ridge waveguides (X-cut lithium niobate), due to the birefringence of the material. This work gives guidelines for designing waveguides in LNOI that do not suffer from the lateral leakage effect. By applying these design considerations, we avoided the lateral leakage effect at the second harmonic wavelength of a nonlinear optical waveguide in LNOI and demonstrate a peak second harmonic generation conversion efficiency of ~1160% $W^{-1}cm^{-2}$.

## 1. Introduction

Lithium niobate on insulator (LNOI) is an emerging photonic integrated circuit (PIC) platform that offers similar integration densities as platforms based on silicon nitride (SiN) and silicon on insulator (SOI), with the advantage of having a strong second order nonlinear optical and electro optical effect [1,2]. Low loss optical waveguides have been achieved in this platform by etching the lithium niobate (LN) directly to form a ridge/rib or wire waveguide [3,4] and by optical loading the LN thin film with a strip of material that has a higher refractive index compared to silica [5-7]. One of the most commonly used optical loading material is SiN due to the wide transparency window and the ease of deposition and patterning. The crystal orientation of the LN thin-film plays an important role in the waveguide design. The most common crystal orientation for the LN thin-film is either X cut or Y cut with the crystallographic Z axis is in the plane of the LN thin-film, with the strongly electro-optic ($r_{33}$) or nonlinear optic ($d_{33}$) tensor components oriented laterally across the waveguides. This orientation enables the use coplanar electrodes placed symmetrically on either side of a waveguide to modulate the waveguide index [8,9] or to use similar electrodes with higher voltage to pole waveguides and achieve quasi-phasematching structures for nonlinear optical applications [5,10]. These properties enabled record breaking demonstrations, such as optical modulation with a voltage length product of 2.8 Vcm [8] and a second-harmonic generation (SHG) efficiency of 2600% $W^{-1}cm^{-2}$ in periodically poled waveguides [10].

The SHG efficiency of 2600% $W^{-1}cm^{-2}$ [10] was achieved by etching the LN directly, which is effective, but requires dedicated fabrication approaches. Furthermore, the quasi-phasematching wavelength is strongly dependent on the exact waveguide dimensions, which makes it intolerant of fabrication errors. A way to avoid these challenges is to form waveguides in un-etched thin films of LN by dielectric loading with a thin strip of moderately high index material. SiN is an attractive choice having sufficiently high index for relatively compact waveguide, but harnessing the mature and CMOS compatible fabrication techniques of the SiN as only the SiN needs to be deposited and etched [11]. We calculated that a waveguide width change of the SiN loaded nonlinear optical waveguides in [5] shifts the quasi-phasematching wavelength by ~0.35 nm for a 10 nm change in the waveguide width. This is 24 times smaller than the quasi-phasematching wavelength shift of the direct-etched waveguides in [10], which shifts by ~8.47 nm for a 10 nm change in the waveguide width. However, the recently demonstrated conversion efficiency using SiN loading on LNOI (160% $W^{-1}cm^{-2}$) was an order of magnitude lower compared to the theoretically predicted one (1600% $W^{-1}cm^{-2}$) [5]. The difference between the experimental and theoretical conversion efficiency has, to date, not been completely understood.

In this contribution, we investigate the impact of lateral leakage on SHG efficiency in SiN strip loaded LNOI and find that the generated second harmonic (SH) can leak into the slab modes due to TE to TM conversion, reducing the overall conversion efficiency of the device. We show how to design optical waveguides to avoid this effect. Applying these design considerations enabled us to demonstrate periodically poled SiN loaded LNOI waveguides with a peak conversion efficiency of ~1160% $W^{-1}cm^{-2}$.

## 2. Simulation and design

Lateral leakage describes an effect where light from a shallow etched waveguide mode radiates into orthogonally polarized slab modes that have higher effective refractive indices. In photonics, this effect was first observed in silicon on insulator (SOI), when the waveguide was excited with a TM mode (see illustration in Fig. 1(a)) [12-15]. In this case, the TM waveguide mode can have a lower effective refractive index compared to the orthogonally polarized TE slab modes, because the TM waveguide mode has a strong evanescent field outside of the waveguide. The angle Θ under which the TM waveguide mode couples to the TE slab modes depends on the phasematching ($n_{TE} \sin(\Theta) = n_{TM}$) [16]. In other words, if a shallow etched SOI ridge waveguide is excited with a TM waveguide mode, then the waveguide will radiate beams of TE light into the slab on each side of the waveguide under an angle Θ. The strength of radiation at each wall is determined by the geometry of that wall [17]. This effect can be seen as a contribution to waveguide loss, which is unwanted in most cases.

In SOI lateral leakage can only occur in ridge waveguides when they are excited with a TM mode, as the orthogonally polarized TE slab modes can have a significantly higher effective refractive index. However, this is not the case in LNOI. Here, the crystal orientation and the orientation of the light polarization needs to be carefully considered, as LN is birefringent. For example, the LN crystal exhibits an extraordinary refractive index ($n_e$) of ~2.138 and an ordinary refractive index ($n_o$) of ~2.211, for a wavelength of 1.55 μm.

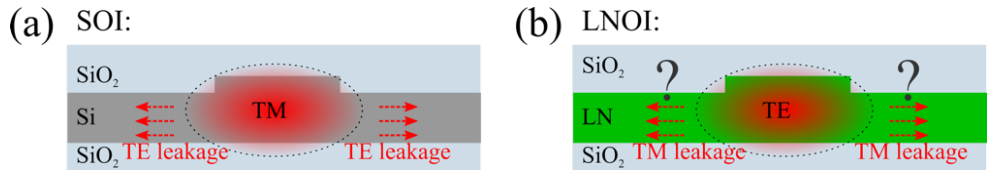

Fig. 1. (a) In SOI lateral leakage of TE slab modes can occur when a ridge waveguides is excited with a TM mode. (b) Can lateral leakage to TM slab modes occur when a LNOI ridge waveguide is excited with a TE mode?

As mentioned in the introduction, it is common to have the polar Z-axis in the plane of the LN thin-film, if one wants to utilize either the strongest electro-optical effect or fabricate periodically poled nonlinear optical waveguide in LNOI by using electrodes on either side of the waveguide. This also requires the excitation of the waveguide with a TE mode, to align the polarization of the light with the polar Z crystal axis (to use either $r_{33}$ or $d_{33}$). To investigate if LNOI can exhibit lateral leakage, we calculated the effective index difference between the TE polarized waveguide mode and the TM polarized slab mode by $\Delta n_{eff} = n_{TE,wg} - n_{TM,slab}$. Lateral leakage can only occur when the effective refractive index of the waveguide mode is lower than the orthogonally polarized slab mode. For negative effective index difference values, one can therefore expect that lateral leakage can take place, which would manifest as additional source of waveguide loss, as illustrated in Fig. 1(b).

*2.1 Lithium niobate etched ridge waveguides in LNOI*

C. Wang et al. [10] demonstrated periodically poled LNOI waveguides with an efficiency of 2600% $W^{-1}cm^{-2}$, which is close to the theoretical efficiency of these waveguides of ~4000% $W^{-1}cm^{-2}$. This indicates that these waveguides did not suffer from lateral leakage at the fundamental and the SH wavelength. To investigate if this is indeed the case, we simulated the difference between the effective index of the guided TE waveguide mode and the radiating TM slab mode for a LNOI ridge waveguide with a width of $w_1 = 1.4$ µm and a etch depth of $h_1 = 300$ nm (similar to C. Wang et al. [10]), as a 2D function of the wavelength and the LN thin-film thickness $h_2$. The results are shown in Fig. 2(a). For convenience we plotted a line in the graph that indicates where the effective index difference is zero. It can be seen that for nearly all wavelengths and $h_2$ the TE waveguide mode has an effective refractive index that is higher than the TM radiating mode with the difference getting higher for smaller $h_2$ and shorter wavelengths. Only in the top left corner (for short wavelengths and large $h_2$) a negative refractive index difference can be observed. These results indicate that for most of the investigated waveguide cross-sections lateral leakage will not take place, this includes both the fundamental and SH wavelengths for the cross-section from C. Wang et al. [10], when using an $h_2$ of 600 nm.

Next, we investigated the effective refractive index difference in the case that the etch depth is reduced to only $h_1 = 150$ nm. This shallower ridge waveguide might be attractive for low-loss waveguides as less field from the optical mode would interact with the waveguide sidewall. Fig. 2(b) shows the difference of effective refractive indices as a function of the wavelength and $h_2$. It can be seen that the curve indicates that a refractive index difference of zero has moved from the top left corner further to the center of the graph. This results in a larger distribution of waveguide cross-sections (for short wavelengths and large $h_2$) that can have TE waveguide modes with effective index lower than the radiating TM slab around them. This indicates that the TE waveguide mode for these parameters can suffer from lateral leakage, which will manifest as an addition loss. If one wants to design nonlinear optical waveguides with such an etch depth, one needs to consider this.

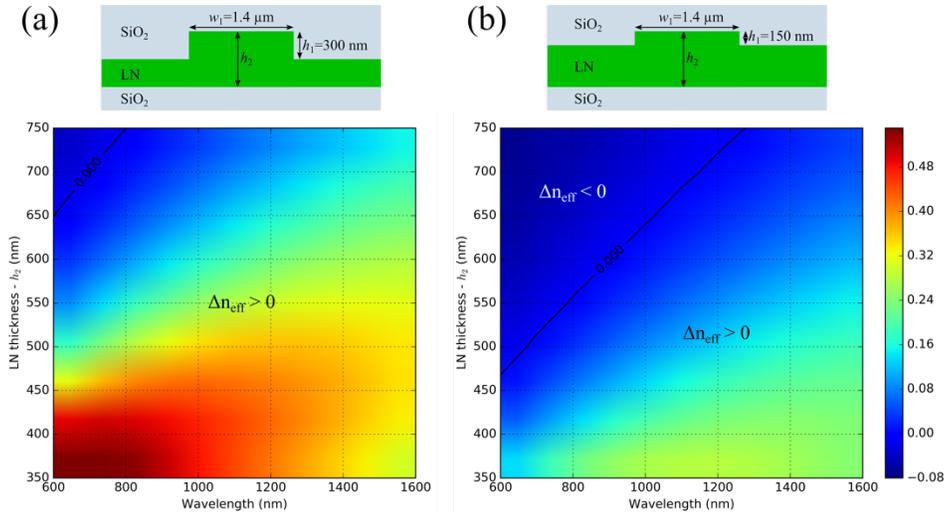

Fig. 2. Effective refractive index difference between the TE waveguide mode and the TM slab mode as a 2D function of the wavelength and LN thickness $h_2$, for an etch depth $h_1$ of 300 nm (a) and 150 nm (b).

To further investigate this, we calculated the effective refractive index difference as a function of the LN etch depth $h_1$ and the LN thickness $h_2$ for the fundamental wavelength 1.55 µm (Fig. 3(a)) and for the second harmonic wavelength 0.775 µm (Fig. 3(b)). To design nonlinear optical waveguides that frequency double a pump wavelength of 1.55 µm, the waveguide dimensions must be selected to be underneath the zero refractive index difference line in Fig. 3(b) in order to avoid any addition losses at the second harmonic wavelength.

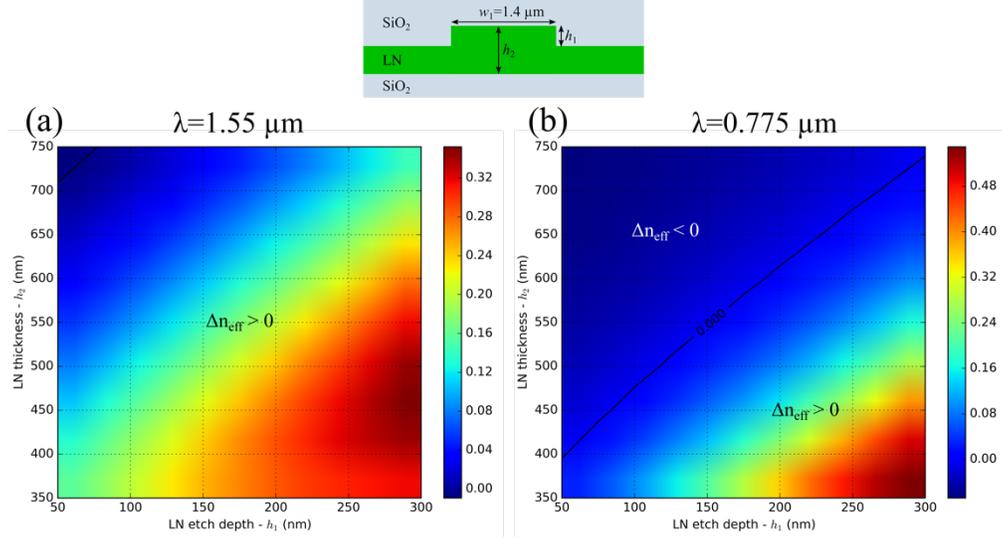

Fig. 3. Effective refractive index difference between the TE waveguide mode and the TM slab mode as a 2D function of the LN etch depth $h_1$ and the LN thickness $h_2$, for a wavelength of 1.55 µm (a) and 0.775 µm (b).

*2.2 Silicon nitride loaded ridge waveguides in LNOI*

L. Chang et al. [5] demonstrated a nonlinear optical conversion efficiency of 160% $W^{-1}cm^{-2}$ for SiN loaded ridge waveguides in LNOI. This value is approximately one order of magnitude lower than the theoretically predicted conversion efficiency. To investigate if these waveguides suffered from lateral leakage we calculated the effective refractive index differences for SiN loaded LNOI ridge waveguides with similar dimensions. We chose a SiN waveguide width of $w_1 = 2$ µm, a SiN thickness of $h_3 = 400$ nm and an etch depth of $h_1 = 380$ nm. The resulting effective refractive index difference between the TE waveguide mode and the TM slab mode is presented in Fig. 4(a) as a function of wavelength and $h_2$. It can be seen that the longest wavelength and largest $h_2$ to exhibit lateral leakage has shifted further into the infrared and larger $h_2$, when compared to the direct etched LN waveguides in Section 2.1. It can also be seen that for an $h_2$ of 700 nm, the SHG wavelength (775 nm) is in the area with a negative refractive index contrast and can therefore suffer from lateral leakage. This indicates that some of the generated SH would leak into the unguided TM slab mode, experiencing an increased loss. When considering that the relationship between measured SHG efficiency and waveguide loss is

$$\frac{P_{2\omega}(L)}{P_\omega(0)} = \eta_o P_\omega(0) L_{\text{eff}} \qquad (1)$$

for a phasematched interaction, where $P_\omega$, $P_{2\omega}$ are the power for the fundamental and second harmonic wavelength, $\eta_o$ is the efficiency for a waveguide that does not suffer from optical losses and

$$L_{\text{eff}} = \frac{\exp(-\alpha_\omega L) - \exp(\alpha_{2\omega} L/2)}{-\alpha_\omega + \alpha_{2\omega}/2} \qquad (2)$$

where $L$ is the nonlinear interaction length, $\alpha_\omega$ and $\alpha_{2\omega}$ are the power loss coefficient for the fundamental and second harmonic wavelength, respectively. This shows that lateral leakage at the second harmonic wavelength effectively reduces the interaction length, which causes a reduction of the measured SHG slope efficiency and could explain the observed lower SHG efficiency in [5].

To reduce the chance of waveguides with lateral leakage, one might consider increasing $h_3$, as this might increase the refractive index difference of the TE waveguide mode to the TM slab mode. To investigate this, we calculated the effective index difference between TE guided and TM radiating modes for an $h_3$ of 600 nm and an $h_1$ of 580 nm. The results are presented in Fig. 4(b). It can be seen that increasing the $h_3$ moves the zero effective refractive index line slightly towards the top left corner of the graph. However, it becomes clear that increasing $h_3$ is beneficial, but not as much as one might anticipate.

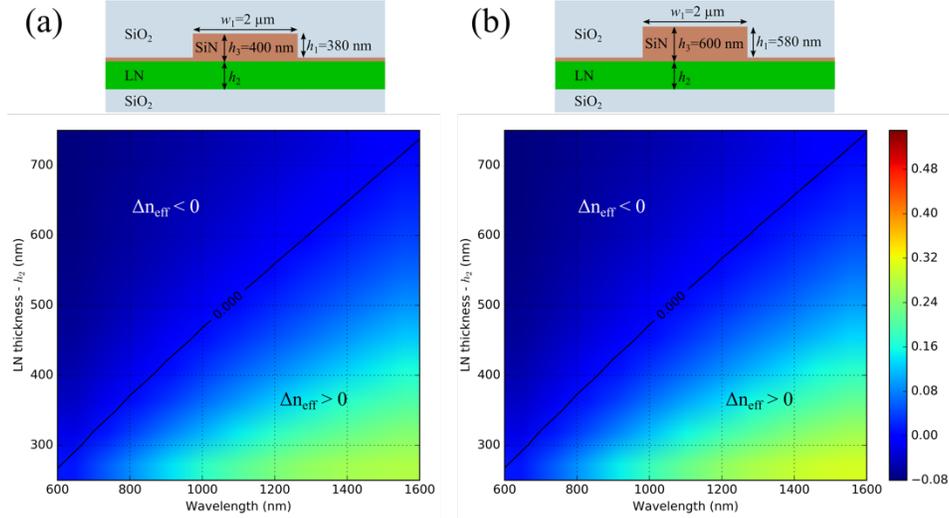

Fig. 4. Effective refractive index difference between the TE waveguide mode and the TM slab mode as a 2D function of the wavelength and the LN thickness. (a) shows the results for an $h_3$ of 400 nm and an $h_1$ of 380 nm and (b) shows the results an $h_3$ of 600 nm and an $h_1$ of 580 nm.

To further investigate this, we calculated the effective refractive index difference as a function of the SiN thickness $h_3$ and the LN thickness $h_2$ for the fundamental wavelength 1.55 μm (Fig. 5(a)) and for the second harmonic wavelength 0.775 μm (Fig. 5(b)). It can be seen that for both wavelengths the LN thickness $h_2$ is dominating the effective index difference. This can be explained by the fact that most of the optical mode is confined in the LN and only a small component is in the SiN loaded area. Increasing the $h_3$ has therefore only a small effect on the overall effective refractive index of the TE waveguide mode.

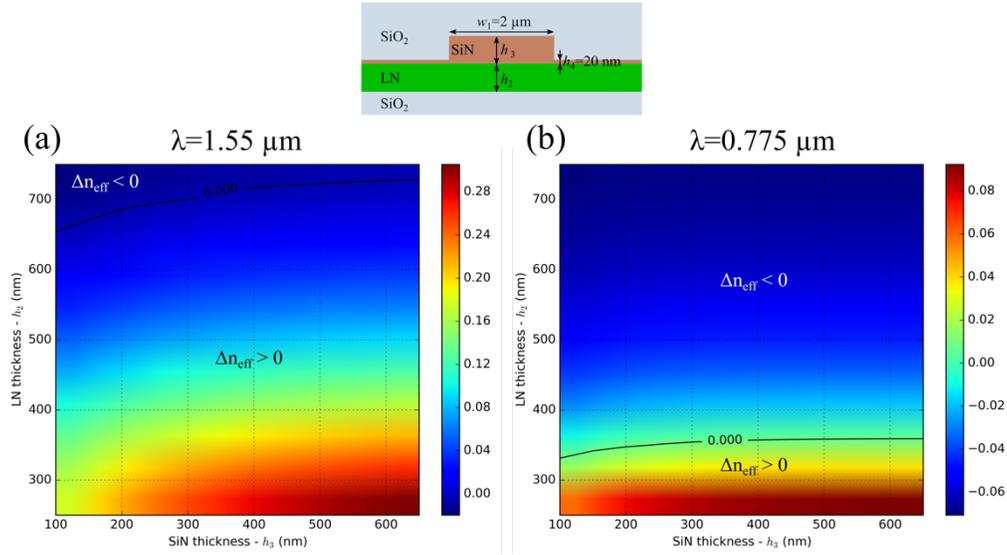

Fig. 5. Effective refractive index difference between the TE waveguide mode and the TM slab mode as a 2D function of the SiN thickness an $h_3$ and the LN thickness an $h_2$. (a) shows the results for a wavelength of 1.55 µm and (b) for a wavelength of 0.775 µm.

## 3. Experimental results and discussion

As a first step, we experimentally investigate lateral leakage in SiN loaded LNOI waveguides that have a similar cross-section to [5]. This investigation is described in Section 3.1. Afterwards (in Section 3.2), we investigate a SiN loaded waveguide cross-section with a 300 nm thick LN thin-film, which should not suffer from lateral leakage at the fundamental and SH wavelength for efficient SHG.

### 3.1 Measurement of lateral leakage loss in silicon nitride loaded ridge waveguide

In Section 2.2 we predicted that the waveguides from [5], which had a LN thickness of 700 nm, may exhibit lateral leakage SH wavelength. However, in Fig. 4(a) one can also notice that the longest wavelength at which lateral leakage can occur is approximately 1.51 µm and therefore quite close to the fundamental wavelength. To investigate if the waveguides used in [5] do indeed exhibit lateral leakage at a wavelength of 1.51 µm, we fabricated an array of similar waveguides and investigated the waveguide transmission as a function of wavelength, between 1.46 µm and 1.578µm. The waveguide length was approximately 5 mm. We used a tunable laser to measure the transmission of the waveguides as a function of wavelength for the TE and the TM mode. Light was coupled in and out of the waveguide by lensed fibers.

The measured transmission spectrum of the waveguide is shown in Fig. 6. It can be seen that both polarizations have periodic oscillations (magnified green spectral area is shown in Fig. 6(b)). The period of the oscillation is approximately 0.11 nm, which matches well with the expected free spectral range of the Fabry-Perot resonance caused by the reflectivity of the waveguide end facets and the cavity length of 5 mm. It can further be observed that the power of the TM mode stays quite constant over the whole wavelength range, however for the TE mode one can observe a sharp drop in transmission for wavelengths below ~1.485 µm (blue highlighted area). The additional loss for these wavelengths is approximately 9 dB/cm. From the simulation results in Fig. 4(a) one would expect lateral leakage to occur at wavelengths of below 1.51 µm and cause additional losses in the waveguide which is higher than the observed wavelength at which increase loss occurs in Fig. 6. This discrepancy may be caused by fabrication tolerances (e.g. the LN thickness was lower than 700 nm) as well as slight differences of the material refractive indices and the assumed Sellmeier equations for the materials [18,19]. For example, a 10 nm thickness variation of the LN thin-film causes a shift of the longest leakage wavelength by ~22 nm. A SiN loaded waveguide with a LN thickness of 690 nm would therefore experience lateral leakage at a wavelength of ~1.488 µm, which is very close to the observed wavelength in the measurements.

Next, we investigated the waveguide behavior for a wavelength of ~780 nm that is close to the SH wavelength for a frequency doubler in this platform. For this we launched the TE and TM mode from an external cavity 780 nm diode laser in the waveguide and measured the transmission. We found that the transmitted power for the TM mode is 16.9 ± 5.8 dB higher than for the TE mode. This indicates that significant leakage takes place in these waveguides at the SH wavelength for the TE mode and may explain the significantly reduced second harmonic efficiency observed with this geometry.

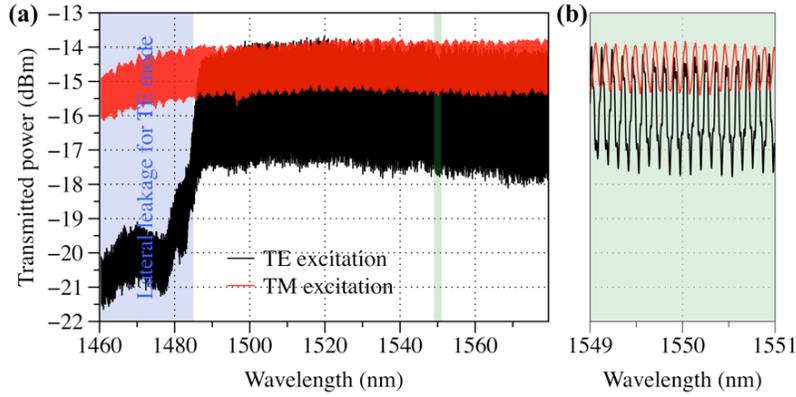

Fig. 6. (a) Transmission of the TE and TM mode as a function of wavelength, for the waveguide cross-section from [5]. (b) Magnified view of the transmission spectrum of the green highlighted area in (a).

*3.2 Avoidance of lateral leakage for efficient second harmonic generation*

According to Fig. 4(a), it should be possible to avoid lateral leakage by reducing the thickness of the LN. If lateral leakage can be eliminated at the second harmonic wavelength, then the efficiency of the second harmonic generation should improve dramatically. To eliminate lateral leakage at a wavelength of 780 nm, we chose a LN thickness of 300 nm (following Fig. 4(a)). We fabricated the waveguides and used the fabrication steps described in [5]. We adjusted the poling period of the domain pattern to 4.98 µm to account for the slightly different effective refractive indices of the fundamental and SH mode. The waveguide length was 4.8 mm. After fabricating the devices, we characterized them using a tunable laser, with a pump power of approximately 1 mW inside of the waveguide. On the output we split the SH and the fundamental wavelength using a wavelength division multiplexer (WDM), before detecting the power with appropriate power meters.

Fig. 7 shows the SHG efficiency as a function of the pump wavelength. One can see strong oscillations in the efficiency, caused by the Fabry-Perot resonances of the fundamental and SH wavelength, which experience a reflection at the end facets of 9.7% and 11.6%, respectively, according to our simulations. The SHG efficiency has an envelope that approximately follows a $sinc^2$ function, which corroborates that the loss at both the fundamental and second harmonic are suppressed, because in the limit where $|\alpha_\omega - \alpha_{2\omega}/2|$ is large, the transfer function should be a broadened Lorentzian (as it was observed in [5]), rather than $sinc^2$.

The bandwidth of the $sinc^2$ envelope is approximately 2.1 nm, which is slightly wider than the theoretical bandwidth of 1.85 nm, predicted using the equation from [20]. The peak conversion efficiency of this waveguide is ~1160% $W^{-1}cm^{-2}$. The theoretical conversion efficiency of the periodically poled LNOI waveguide is 1070% $W^{-1}cm^{-2}$. However, the Fabry-

Perot resonance can increase the fundamental power in a cavity by a factor of $(1-R_\omega)^2$ [21], where $R_\omega$ is the reflectivity of the waveguide end facets. This corresponds to an increase the fundamental power by 22.5% inside the cavity and results in an increased theoretical peak conversion efficiency of 1606% $W^{-1}cm^{-2}$ due to square power dependency.

The lower peak conversion efficiency of the fabricated nonlinear optical waveguides as well as the wider bandwidth can be explained by not considering the waveguide losses, which we estimate to be 0.6 dB/cm for the fundamental wavelength. If one assumes a similar optical loss for the second harmonic wavelength the effective device length using Eq. 2 would be 4.57 mm, which would result in a drop of the conversion efficiency to ~1450% $W^{-1}cm^{-2}$. The lower measured conversion efficiency may be caused by a slightly higher optical loss of the second harmonic wavelength as well as not having a perfect duty cycle of the poling pattern. Nevertheless, the nonlinear optical performance of the waveguide shows that if one avoids lateral leakage at the SH wavelength that one can achieve a high nonlinear optical conversion efficiency in LNOI ridge/rib waveguides.

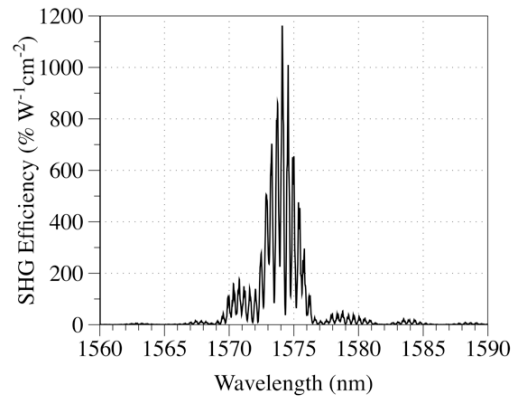

Fig. 7. SHG efficiency of the periodically poled LNOI waveguide with a LN thickness of 300 nm and a SiN thickness of 400 nm.

## 4. Conclusion

We have shown that lateral leakage can take place not only for TM waveguide modes in SOI, but also for TE waveguide modes in LNOI. The occurrence of lateral leakage for a TE waveguide mode can be explained by the birefringence of LN. Hence, lateral leakage must be considered for linear and nonlinear optical ridge/rib waveguides in LNOI as it can cause undesired waveguide losses. Lateral leakage in LNOI ridge/rib waveguides is more likely for shorter wavelengths and larger LN thicknesses for a given ridge/rib height.

We also discovered that the nonlinear optical waveguides in [5] suffered from lateral leakage at the SH wavelength. This helped explain the significantly lower conversion efficiency that was observed in these waveguides than what was expected. By choosing an appropriate waveguide cross-section we avoided lateral leakage in SiN loaded LNOI waveguides, which enabled us to demonstrate a peak SH efficiency of ~1160% $W^{-1}cm^{-2}$.

**Funding**

This research is supported by the ARC DP190102773 and DP190101576 as well as DARPA MTO DODOS contract (HR0011-15-C-055).

**Acknowledgments**

The authors acknowledge the facilities, and the scientific and technical assistance, of the Micro Nano Research Facility (MNRF) and the Australian Microscopy & Microanalysis Research Facility at RMIT University.